\documentclass[a4paper]{jpconf}
\usepackage{graphicx}
\begin{document}

\newcommand{\ctan}{$^{13}$C($\alpha$,n)$^{16}$O }
\newcommand{\nean}{$^{22}$Ne($\alpha$,n)$^{25}$Mg }
\newcommand{\ct}{$^{13}$C }

\title{The FRUITY database on AGB stars: past, present and future}

\author{Sergio Cristallo}
\address{INAF Osservatorio Astronomico di Teramo, via Maggini sn 64100 - Teramo (Italy)}
\address{INFN Sezione Napoli, Napoli (Italy)}

\ead{cristallo@oa-teramo.inaf.it}

\author{Luciano Piersanti}
\address{INAF Osservatorio Astronomico di Teramo, via Maggini sn 64100 - Teramo (Italy)}
\address{INFN Sezione Napoli, Napoli (Italy)}

\author{Oscar Straniero}
\address{INAF Osservatorio Astronomico di Teramo, via Maggini sn 64100 - Teramo (Italy)}
\address{INFN Sezione Napoli, Napoli (Italy)}

\begin{abstract}
We present and show the features of the FRUITY database, an
interactive web-based interface devoted to the nucleosynthesis in
AGB stars. We describe the current available set of AGB models
(largely expanded with respect to the original one) with masses in
the range 1.3$\le $M/M$_\odot \le$3.0 and metallicities
$-2.15\le$[Fe/H]$\le$+0.15. We illustrate the details of our
s-process surface distributions and we compare our results to
observations. Moreover, we introduce a new set of models where the
effects of rotation are taken into account. Finally, we shortly
describe next planned upgrades.
\end{abstract}

\section{Introduction}

Asymptotic Giant Branch stars (hereafter AGBs) are exceptional
laboratories to test the robustness of stellar models. During this
evolutionary phase, both light (C, N, F, Na) and heavy elements
can be produced. The latter are synthesized via the slow neutron
capture process (the s-process). A detail description of AGBs
evolution and nucleosynthesis
can be found in [1].\\
The structure of an AGB consists of a partially degenerate CO
core, surrounded by an He-shell and an H-shell separated by a thin
layer (He-intershell) and by a cool and expanded convective
envelope. The surface luminosity is mainly sustained by the
H-burning shell, active for most of the time. This situation is
recurrently interrupted by the growing up of thermonuclear
runaways driven by violent ignitions of 3$\alpha$ reactions at the
base of the He-intershell (Thermal Pulses, hereafter TPs). The
energy suddenly released by TPs can not be transported outward
radiatively and, thus, convective episodes develop. These
convective shells efficiently mix the He-intershell, enriching
this layer in carbon (produced by 3$\alpha$ reactions) and in
s-process elements. Moreover, the energy boosted by the TP forces
the overlying layers to expand, possibly switching off the
H-shell. This allows the convective envelope to penetrate
downward, carrying to the surface the isotopes freshly synthesized
in the He-intershell. This phenomenon is called Third Dredge Up
(hereafter TDU) and it is know to work in AGB stars since the
early fifties [2]. \\
The major neutron source in AGB stars is the \ctan reaction.
Neutrons are released within the He-intershell during the
interpulse phase (thus in radiative conditions) when T$\sim
1.0\times 10^8$ K [3]. Actually, the amount of \ct requested to
fit observations is one of the major sources of uncertainty   in
AGB models ([4]; see also Section \ref{modelli}). A marginal
contribution comes from the \nean reaction, which is activated at
higher temperatures (T$> 2.5\times 10^8$ K) during TPs.
\begin{figure}[t]
\includegraphics[width=38pc]{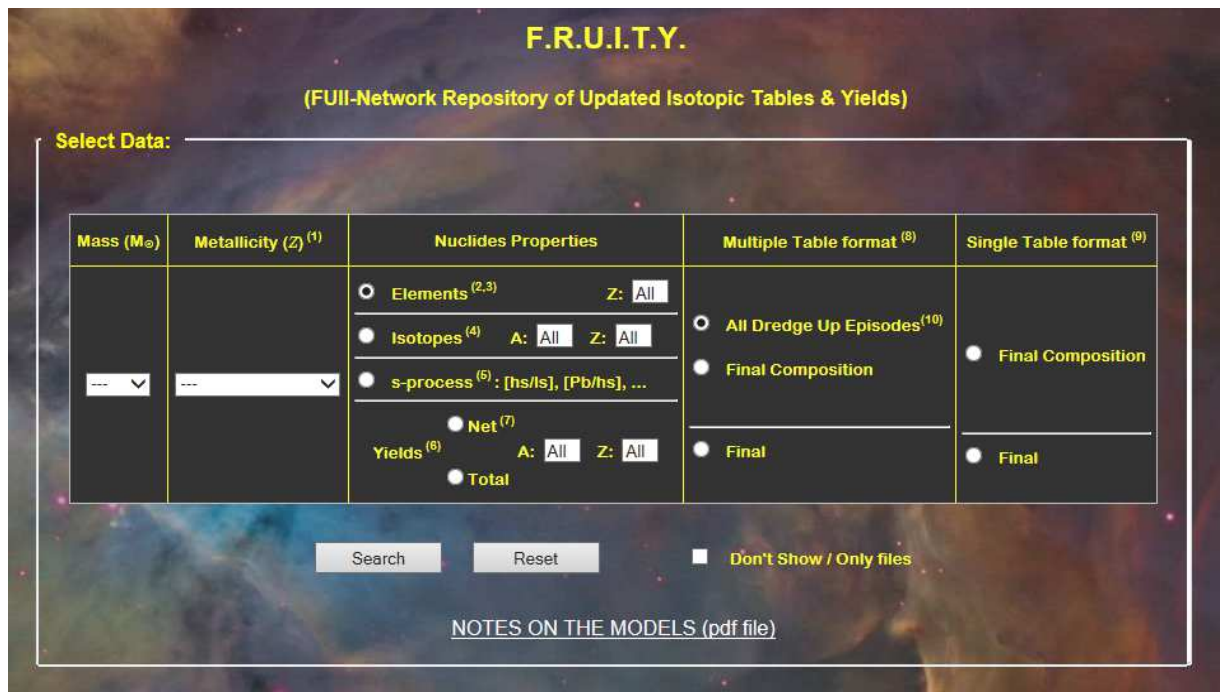}
\caption{The FRUITY database homepage.} \label{fig0}
\end{figure}

The great amount of available spectroscopic data such as the need
of yields for Chemical Evolution Models require the computation of
a large number of detailed AGB evolutionary models. We fit this
request by making available our theoretical results on the on-line
web pages of the FRUITY (FUNs Repository of Updated Isotopic
Tables \& Yields) database, entirely dedicated to the
nucleosynthesis in AGB stars.

In Section \ref{modelli} we briefly present our evolutionary code
(FUNs), while in Section \ref{database} we describe the set of
models currently available on FRUITY. In Section \ref{rota} we
present a recently published set of rotating AGB stars. Finally,
in Section \ref{up} we list our planned upgrades.

\section{The FUNs evolutionary code} \label{modelli}

The FUNs (FUll Network stellar) evolutionary code is a
one-dimension hydrostatic code (see [1] and references therein).
The adopted mass-loss rate has been calibrated on the
Period-Luminosity and Period- Mass loss relations observed in Long
Period Variable Stars [1]. In the envelope, atomic and molecular
opacities are calculated according to the changes in the chemical
composition due to the occurrence of TDU episodes [4]. The
radiation/convection interface at the inner border of the
convective envelope is treated by applying an exponential decay of
the convective velocities. As a by-product, we obtain the
self-consistent formation of the \ct pocket after each TP followed
by TDU. The extension of the that pocket varies from TP to TP
following the shrinking of the He-intershell [5]. In order to
avoid the possible loss of accuracy inherent to post-process
techniques, we directly coupled our models to a full nuclear
network, which includes all isotopes from hydrogen to bismuth (at
the ending-point of the s-process path).

\section{The FRUITY database} \label{database}

The original set of FRUITY has been presented in [6]. It consists
of 28 evolutionary models, with different combinations of masses
(1.5, 2.0, 2.5, and 3.0 M$_\odot$) and metallicities ([Fe/H]=
-1.15, -0.67, -0.37, -0.24, -0.15, 0.00, +0.15). Different He
contents and scaled solar compositions [7] are used (see [6] for
details). \\
Recently, new models have been uploaded. In particular, the mass
range has been expanded down to 1.3 M$_\odot$ models. Note that,
at large metallicities, these masses do not experience TDU
episodes. \\
In order to cover the metallicity range of Galactic Globular
Clusters, we compute two additional metallicities: [Fe/H]=-2.15
and [Fe/H]=-1.67, both of them  with $\alpha$-enhanced
elements ([$\alpha$/Fe]=0.5).\\
For each model, users can freely download pulse by pulse surface
isotopic compositions and elemental overabundances\footnote{In the
usual spectroscopic notation: [El/Fe] = $log$(N(El)/N(Fe))$_*$ -
$log$(N(El)/N(Fe))$_\odot$}. Tables are available in two different
formats (see Figure \ref{fig0}). In the Multiple Table Format, the
query returns multiple table, depending on the number of selected
models. In the Single Table Format, the query returns a single
table containing all the selected models; in this case, only
single elements (or isotopes) can be visualized. Among other
quantities available on FRUITY there are the
[ls/Fe]\footnote{[ls/Fe] = ([Sr/Fe]+[Y/Fe]+[Zr/Fe])/3}, the
[hs/Fe]\footnote{[hs/Fe] = ([Ba/Fe]+[La/Fe]+[Nd/Fe]+[Sm/Fe])/4}
and the s-process indexes [hs/ls], [Pb/hs] and [Pb/ls]. Finally,
stellar yields can be selected. With respect to the original
version of the FRUITY database, beside net yields we also made
available total stellar yields.\\
Users interested in upgrades of the FRUITY database can register
to the dedicated mailing list.

\section{Comparison to observations} \label{compa}

In order to verify the robustness of our models, we compare them
to their observational counterparts. In particular, we concentrate
on the Luminosity Function of Carbon Stars (hereafter LFCS) and on
{\it s}-process spectroscopic indexes at different metallicities.

The LFCS links the luminosity of these objects (and thus the core
mass; see [8]) with their carbon surface abundance. Thus, by
studying this quantity, we can test our prescriptions on
convection and mass-loss law. In 2006, [9] analyzed a sample of
Galactic Carbon Stars and show how the radiation emitted by these
objects at infrared wavelengths strongly weights the LFCS toward
large lambdas. Recently, [10] proposed a re-analysis of the same
sample and found that the high luminosity tail of the LFCS
described by [9] disappears, in good agreement with theoretical
expectations [6].
\begin{figure}[t]
\includegraphics[width=34pc]{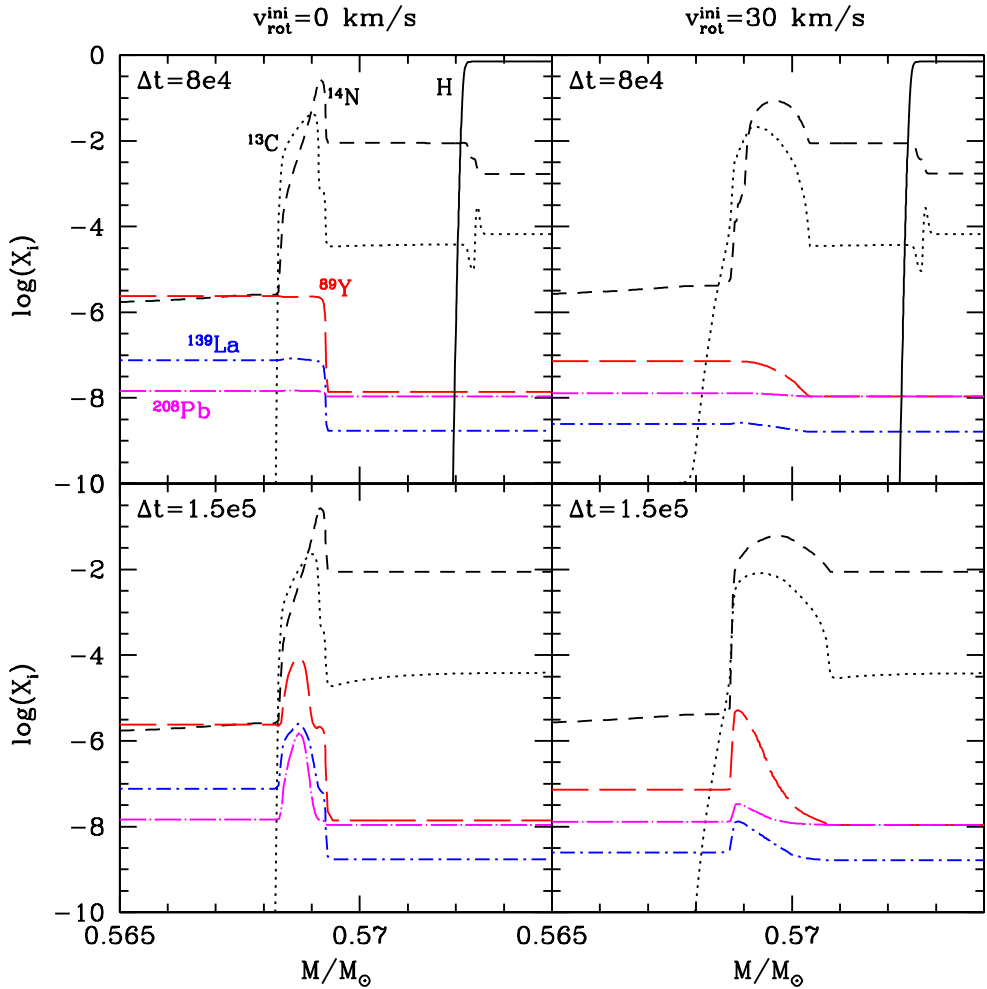}
\caption{Temporal evolution of key elements during the radiative
burning of the $2^{nd}$ $^{13}$C pocket of a 2 M$_\odot$ star with
Z=$10^{-2}$ and different initial rotation velocities. See text
for details.} \label{fig1} \end{figure}

Another useful litmus test for theoretical models is the study of
s-process spectroscopic indexes, in particular the [hs/ls] and the
[Pb/hs]. Before introducing them, a distinction between intrinsic
and extrinsic s-process rich stars has to be
done.\\
TP-AGB stars are intrinsic s-rich stars, since they are presently
undergoing thermal pulses and TDU episodes. Post-AGB
stars belong to the same class. \\
Extrinsic s-rich stars are less-evolved stars (dwarfs or giants)
belonging to binary systems, whose s-enhancement is likely due to
the pollution caused by the intense wind of an AGB companion.
Ba-stars and, at lower metallicities, CH stars and Carbon Enhanced
Metal Poor s-process rich stars (hereafter CEMP-s stars)
belong to this class of objects.\\
The only possibility to group together all s-rich stars is to
study relative surface abundances. In such a way, any problem
correlated to possible further dilution process or to their
evolutionary status is avoided.\\
We have already shown that our theoretical models match
observations, even if they are not able to explain the observed
spread for a fixed metallicity (see Fig. 12 and 13 of [6]).
Although it is possible that (at least part of) the observed
spread can be ascribed to the large observational uncertainties,
theoretical scenarios able to explain such a spread has to be
explored. Among them, we verified is the mixing induces by
rotation can lead to a certain spread of the spectroscopic
indexes.
\begin{figure}[t]
\includegraphics[width=34pc]{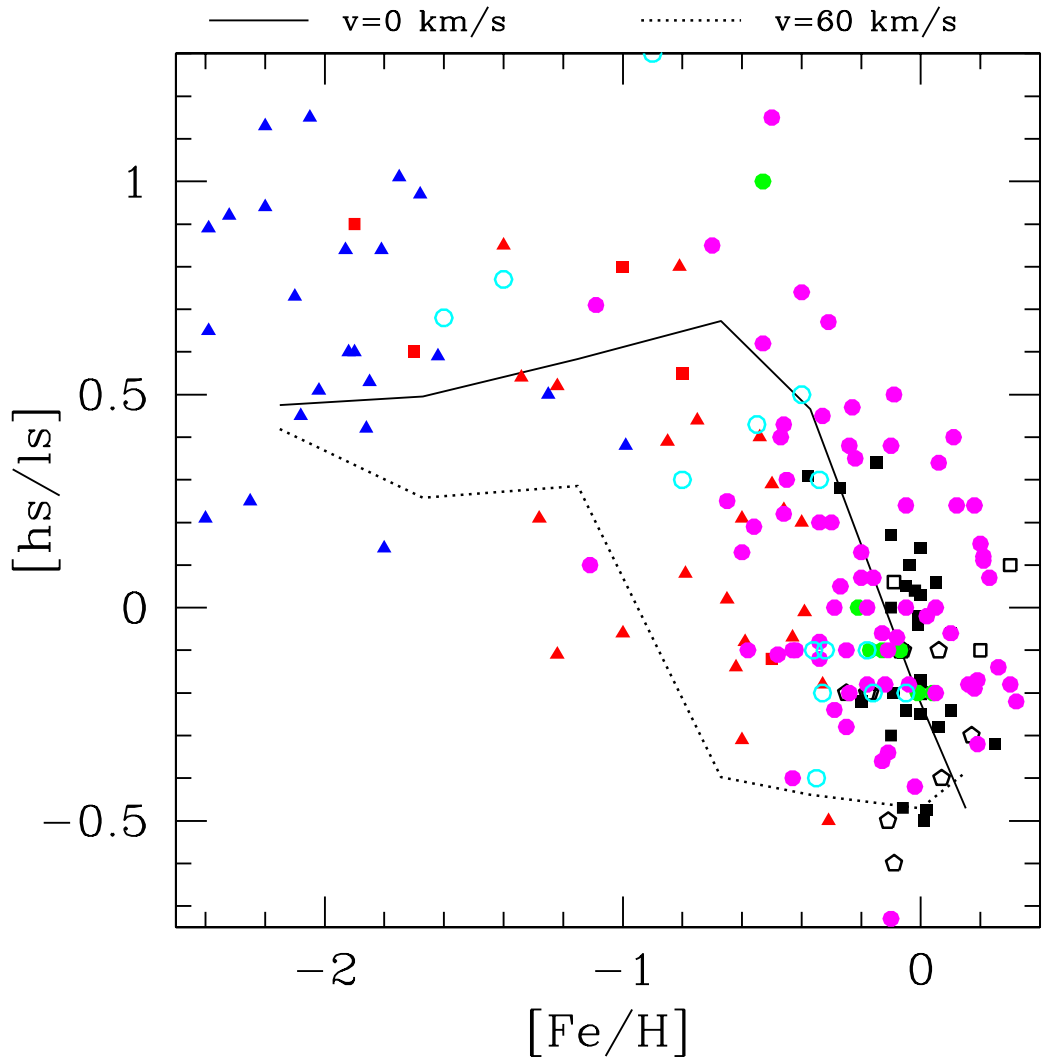}
\caption{[hs/ls] s-process index as a function of metallicity for
models with and without rotation. See text for details.}
\label{fig2}
\end{figure}

\section{Low mass rotating AGB models} \label{rota}

\begin{figure}[t]
\includegraphics[width=34pc]{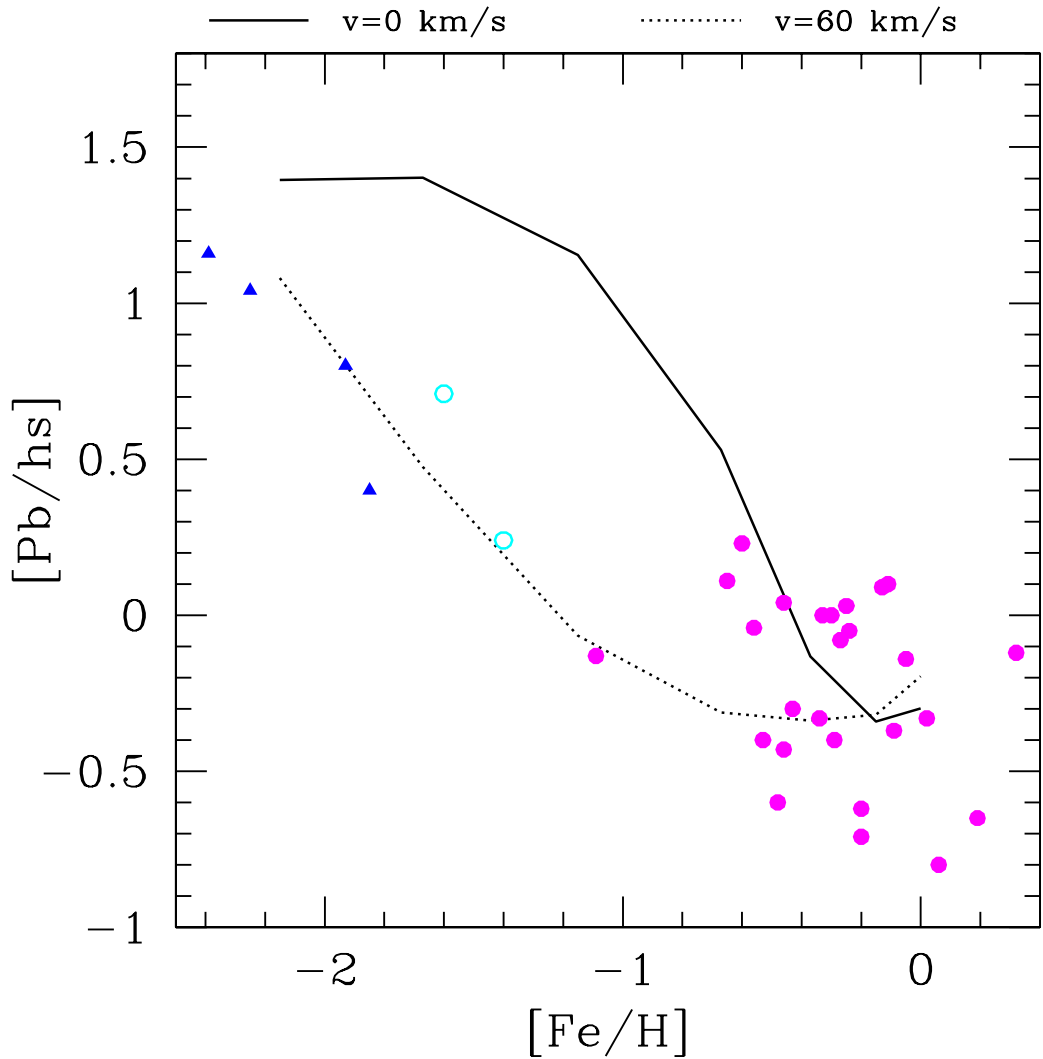}
\caption{As in Figure \ref{fig2}, but for the [Pb/hs] s-process
index.} \label{fig3} \end{figure}

Even if low mass stars are generally slow rotators, the lifting
due to the centrifugal force and the mixing induced by dynamical
and secular instabilities can modify their physical structure and
chemical composition. In order to verify the effect that rotation
can have on the nucleosynthesis during the AGB phase, we implement
our models with rotation. In particular, we add rotation-induced
mixing to the other mixing already considered in our previous
works, those due to convection in particular. We found that a
variation in the initial velocity can lead to a consistent spread
in the final surface s-process enhancements and spectroscopic
indexes for stars with the same
initial mass and metallicity [11]. \\
Rotation does not determine the formation of the $^{13}$C pocket.
Notwithstanding, rotation-induced mixing modify the mass extension
of both the $^{13}$C and the $^{14}$N pockets and their overlap,
thus reducing the average neutrons-to-seeds ratio. In Figure
\ref{fig1} we report the $^{13}$C and the $^{14}$N abundances in
the region where the $2^{nd}$ $^{13}$C pocket of a 2 M$_\odot$
star with Z=$10^{-2}$  forms. Left panels refer to a non-rotating
model, while right panels to a model with initial rotation
velocity $v^{ini}_{rot}$=30 km/s~. In the same plot we also report
the local abundances of $^{89}$Y, $^{139}$La and $^{208}$Pb,
assumed as representative of the three s-process peaks. We find
that the Goldreich-Schubert-Fricke instability, active at the
interface between the convective envelope and the rapid rotating
core, contaminates the $^{13}$C-pocket with $^{14}$N. Thus,the
mixing induced by rotation locally decreases the neutron-to-seed
ratio, leading to a reduction of the total amount of heavy
elements produced by the s process and favoring the light-s
elements with respect to the heavier ones. As a matter of fact,
both the [hs/ls] and the [Pb/hs] spectroscopic indexes decrease as
the initial rotation velocity increases (see Figures 6 and 9 of
[11]). At low metallicity, the combined effect of
Goldreich-Schubert-Fricke instability and meridional circulations
determines an increase of light-s and, to a less extent, heavy-s
elements, while lead is strongly reduced. \\
In Figures \ref{fig2} and \ref{fig3} we report, as a function of
the metallicity, the [hs/ls] and [Pb/hs] s-process indexes for
models with and without rotation. Two values of initial rotation
velocities ($v^{ini}_{rot}$=60 km/s and $v^{ini}_{rot}$=120 km/s)
have been considered. With respect to the observational data
reported in [6], we add recent measurements of post-AGB stars
[12,13,14,15], Ba-stars [16,17], CH stars [18,19,20] and a
selection of well measured CEMP-s stars (see [21] and references
therein).

Our results suggest that rotation can be regarded as a possible
physical mechanism responsible for the observed spread of
s-process spectroscopic indexes. In particular, the large spread
in the [hs/ls] at intermediate metallicities (-1.0$<$[Fe/H]$<
$-0.6) can be reproduced by hypothesizing different initial
rotation velocities. Notwithstanding, our models cannot match the
large [hs/ls] (up to 1) measured at low metallicities.
Interestingly, the lower [Pb/hs] values characterizing rotating
models are in agreement with observations. In fact, as shown in
Figure \ref{fig3}, models with $v^{ini}_{rot}$=60 km/s better
reproduce the sequence, starting from high Z, of Ba-stars (full
dots), CH stars (open dots) and CEMP-s stars (full triangles).
Unfortunately, firm conclusions cannot be drawn due to the paucity
of lead measurements and to the large errors affecting
observations.

\section{Planned Upgrades} \label{up}

The current version of FRUITY spans on a reasonable range of
metallicities and well covers the low mass range. Intermediate
mass AGB stars (4$<$ M/M$_\odot <$ 7; hereafter IM-AGBs) are
currently missing. These stars are particulary important for the
chemical evolution of Globular Clusters (see e.g. [22] and
references therein). We already conducted an explorative study of
IM-AGBs [23], but without publishing detailed nucleosynthetic
predictions. Therefore, we intend to explore the evolution and
nucleosynthesis of IM-AGBs, starting from the metallicities of
interest for the study of Galactic Globular Clusters [24]. Later,
we aim to extend these calculations to all
the metallicities considered in FRUITY.\\
Finally, as a long-term project, we are planning to extend the
mass range of FRUITY to massive stars, in order to study the weak
component of the s-process, at work during the core-He burning and
the C-burning shell phases particularly efficient when rotation is
explicitly taken into account. (see [25] and references therein).

\subsection{Acknowledgments} The authors warmly thank Prof. Roberto
Gallino for continuous and fruitful scientific discussions.

\end{document}